\documentclass[12pt]{article}
\usepackage{graphicx,amsmath,amssymb,color,bm}

\begin{document}

\centerline {\Large\textbf {Novel Magnetic Quantization of sp$^{3}$ Bonding in
Monolayer Tinene}}
%\centerline {\Large\textbf {Novel Magnetic Quantization of sp$^{3}$ Bonding in
%Monolayer Tin}}

%\centerline {\Large \textbf {}}\vskip0.6 truecm

\centerline{Szu-Chao Chen$^{1}$, Chung-Lin Wu$^{2}$, Jhao-Ying Wu$^{2,\star}$, and Ming-Fa
Lin$^{2,\star}$ }
\centerline{$^{1}$Center for Micro/Nano Science and
Technology, National Cheng Kung University, Tainan, Taiwan 701}
\centerline{$^{2}$Department of Physics, National Cheng Kung University,
Tainan, Taiwan 701} \vskip0.6 truecm

\noindent A generalized tight-binding model, which is based on the
subenvelope functions of the different sublattices, is developed to explore
the novel magnetic quantization in monolayer gray tin. The effects due to
the $sp^{3}$ bonding, the spin-orbital coupling, the magnetic field and the
electric field are simultaneously taken into consideration. The unique
magneto-electronic properties lie in two groups of low-lying Landau levels,
with different orbital components, localization centers, state degeneracy,
spin configurations, and magnetic- and electric-field dependences. The first
and second groups mainly come from the $5p_{z}$ and ($5p_{x}$,$5p_{y}$)
orbitals, respectively. Their Landau-level splittings are, respectively,
induced by the electric field and spin-orbital interactions. The intragroup
anti-crossings are only revealed in the former. The unique tinene Landau
levels are absent in graphene, silicene and germanene.

\vskip0.6 truecm

$\mathit{PACS}$:

\newpage

\bigskip

\centerline {\textbf {I. INTRODUCTION}} \bigskip The layered group-IV
condensed-matter systems have attracted enormous attention,$^{1-9}$ mainly
owing to their nano-scaled thickness and hexagonal symmetry. They are ideal
two-dimensional (2D) materials for studying novel physical, chemical and
material phenomena,$^{1-6,10-16}$ and they have high potential for technical
applications$^{4,7,17,18}$ in the near future. Graphene, with the strong $%
sp^{2}$ bonding in the one-atom-thick plane, was the first stable 2D crystal
produced by mechanical exfoliation in 2004.$^{1}$ Its Si and Ge
counterparters, silicene and germanene, have been synthesized on metallic
substrates.$^{7,8,19-22}$ Recently, a single layer of Sn atoms, tinene, has
been fabricated on a substrate of bismuth telluride.$^{9}$ Silicene,
germanene and tinene have low-buckled structures (Fig. 1(b)) with a mixed $%
sp^{2}$-$sp^{3}$ hybridization rather than a $sp^{2}$ one, since the
relatively weak chemical bonding between the larger atoms cannot maintain a
planar structure. Furthermore, the spin-orbital coupling (SOC) is evidently
stronger than that of graphene and can open an energy gap. The effects of
the $sp^{3}$ bonding and SOC are pronounced in silicene and germanene and
are expected to become even more significant in tinene. The magnetic
quantization in tinene will be investigated by using the generalized
tight-binding model,$^{23-25}$ in which all the critical interactions are
taken into account simultaneously. Comparison with the other 2D systems is
also made.

The hexagonal symmetry causes graphene to exhibit a pair of linear ($\pi
,\pi ^{\ast }$) bands intersecting at the Fermi level ($E_{F}=0$) near the
corners of the hexagonal Brillouin zone (K and K$^{\prime }$ in the inset of
Fig. 2(a)). However, the SOC slightly separates two Dirac points and
modifies the energy dispersions in silicene and germanene. Their low-energy $%
\pi $-electronic structures could be well described by the effective-mass
model,$^{26}$ being consistent with the first-principles calculations.$^{14}$
On the other hand, tinene has the strong $sp^{3}$ bonding and SOC, leading
to the low-lying $\pi $ and $\sigma $\ bands formed by the $5p_{z}$ and ($%
5p_{x}$,$5p_{y}$) orbitals, respectively. These special low-energy bands are
expected to enrich the magnetic quantization.

A uniform perpendicular magnetic field ($\mathbf{B}_{z}\mathbf{=}B_{z}$ $%
\widehat{z}$) makes electronic states flock together and induces many Landau
levels (LLs). For graphene, silicene and germanene, the low-lying $\pi $%
-electronic states can be directly quantized by the effective-mass model.
The LL energies are approximated by a square-root form $E^{c,v}=sgn(n^{c,v})%
\sqrt{(-\frac{1}{2}\xi \eta \Delta _{so}^{{}})^{2}+2\left\vert
n^{c,v}\right\vert v_{F}^{2}\hbar eB_{z}/c}$, where $c$ represents the
conduction state, $v$ the valence state, $n^{c,v}$ the LL quantum mode, $\xi
=\pm 1$ the K or K$^{^{\prime }}$\ valley index, $\eta =\pm 1$ the spin
index, $\Delta _{SO}^{{}}$ the spin-orbital band gap, and $v_{F}^{{}}$ the
Fermi velocity.$^{10-12}$ The $\sqrt{B_{z}}$-dependent LL energies in
graphene with $\Delta _{SO}^{{}}=0$ have been confirmed by scanning
tunneling spectroscopy (STS)$^{27,28}$ and magneto-optical experiments of
cyclotron resonance.$^{29}$ Specially, the stronger SOC$^{^{\prime }}$s in
silicene and germanene lead to the splitting of the $n^{c,v}=0$ LL with
energy spacing $\Delta _{so}^{{}}$.$^{11-13}$

The effective-mass model may be too cumbersome or complex to study the
magneto-electronic properties of monolayer tinene with its strong $sp^{3}$
bonding and SOC, although it could conceivably be used to comprehend the
low-energy magnetic quantization of graphene, silicene and germanene.
Apparently, the feature-rich energy bands of tinene, with multi-valley
structures near $E_{F}$,$^{15}$ cause huge difficulties in directly
diagonalizing the magnetic Hamiltonian matrix. We will develop the
generalized tight-binding model for various orbital hybridizations, spin
configurations, and external fields, in which the Hamiltonian matrix is
built from the orbital- and spin-dependent tight-binding basis functions.
The spatial distributions of the subenvelope functions on the distinct
sublattices are critical for distinguishing the main features of LLs. This
study shows that monolayer tinene exhibits a novel magnetic quantization.
Two groups of LLs, which are dominated by the $5p_{z}$ and ($5p_{x},5p_{y}$)
orbitals, repectively, have distinct localization centers, orbital
components, spin configurations, state degeneracies, and magnetic- and
electric-field dependences. The predicted LL energy spectra can be verified
by STS measurements.$^{30}$

\bigskip

\bigskip \bigskip \centerline {\textbf {II. METHODS}}\bigskip \bigskip

The generalized tight-binding model is further developed to include the $%
sp^{3}$orbital bonding, the SOC, and the electric and magnetic fields
simultaneously. Tinene consists of two equivalent A and B sublattices,
respectively, located at two parallel planes with a separation of l$_{z}$ ($%
0.417$ $\mathring{A}$). The primitive unit vectors are $\mathbf{a}_{1}$ and $%
\mathbf{a}_{2}$ with a lattice constant of a=$4.7$ $\mathring{A}$ (Fig.
1(a)), and the angle between the Sn-Sn bond and the $z$-axis is $\theta
=107.1^{%
%TCIMACRO{\U{b0}}%
%BeginExpansion
{{}^\circ}%
%EndExpansion
}$ (Fig. 1(b)). The $5s$ orbital energy is $E_{s}=-6.2335$ eV below that of
the $5p$ orbitals taken as zero ($E_{p}=0$). The Slater--Koster hopping
parameters in the $sp^{3}$ bonding are $V_{ss\sigma }=-2.6245$ eV, $%
V_{sp\sigma }=2.6504$ eV, $V_{pp\sigma }=1.4926$ eV, and $V_{pp\pi }=-0.7877$
eV (Fig. 1(c)). The SOC strength ($\lambda _{soc}=0.8$ eV) of tinene is
predicted to be two orders of magnitude greater than that of graphene.$%
^{31,32}$ This interaction under the Coulomb central potential, $%
V_{soc}=\lambda _{soc}\overrightarrow{L}\cdot \overrightarrow{s}$, will
induce dramatic changes in electronic properties. Here, $V_{soc}$
corresponds to the same atom and only the $5p$ orbitals contribute to $%
V_{soc}$. In the bases of $\left\{ \left\vert 5p_{z}^{A}\right\rangle
,\left\vert 5p_{x}^{A}\right\rangle ,\left\vert 5p_{y}^{A}\right\rangle
,\left\vert 5s^{A}\right\rangle ,\left\vert 5p_{z}^{B}\right\rangle
,\left\vert 5p_{x}^{B}\right\rangle ,\left\vert 5p_{y}^{B}\right\rangle
,\left\vert 5s^{B}\right\rangle \right\} \otimes \left\{ \uparrow
,\downarrow \right\} ,$ the nearest-neighbor Hamiltonian is expressed as

$H=\underset{\left\langle i\right\rangle ,o,m}{\sum }%
E_{o}C_{iom}^{+}C_{iom}^{{}}+\underset{\left\langle i,j\right\rangle
,o,o^{\prime },m}{\sum }\gamma _{oo^{\prime }}^{\mathbf{R}%
_{ij}}C_{iom}^{+}C_{jo^{\prime }m}^{{}}+\underset{\left\langle
i\right\rangle ,p_{\alpha },p_{\beta }^{{}},m,m^{\prime }}{\sum }\frac{%
\lambda _{SOC}}{2}C_{ip_{\alpha }m}^{+}C_{ip_{\beta }m^{\prime
}}^{{}}(-i\epsilon _{\alpha \beta \gamma }\sigma _{mm^{\prime }}^{\gamma }),%
\newline
$where $i(j)$, $o(o^{\prime })$, and $m(m^{\prime })$ stand for the lattice
site, atomic orbital, and spin, respectively. The first and second terms
are, respectively, the site energy ($E_{o}$) and the nearest-neighbor
hopping integral ($\gamma _{oo^{\prime }}^{\mathbf{R}_{ij}}$). The latter
depends on the type of atomic orbitals, the translation vector of the
nearest-neighbor atom ($\mathbf{R}_{ij}$), and $\theta $. The sufficiently
strong $sp^{3}$ hybridization not only gives rise to the hoppings between $%
5p_{z}^{{}}$ and $(5p_{x},5p_{y},5s)$ orbitals, e.g., $\gamma _{zx}^{\mathbf{%
R}_{1}}=(V_{pp\pi }-V_{pp\sigma })\sin \theta \cos \theta $, but also leads
to the misorientation of $p\pi $ orbitals ($\gamma _{zz}^{\mathbf{R}%
_{ij}}=V_{pp\pi }\sin ^{2}\theta +V_{pp\sigma }\cos ^{2}\theta $).$^{33}$
For the planar case ($\theta =90^{%
%TCIMACRO{\U{b0}}%
%BeginExpansion
{{}^\circ}%
%EndExpansion
}$), $\left\vert 5p_{z}^{{}}\right\rangle $\ is orthogonal to ($\left\vert
5p_{x}^{{}}\right\rangle $, $\left\vert 5p_{y}^{{}}\right\rangle $, $%
\left\vert 5s\right\rangle )$ and there are only $p\pi $ bondings in
parallel $5p_{z}$ orbitals. The last term represents the $V_{soc}$ on the
same atom where $\alpha ,\beta $ and $\gamma $, respectively, denote the $x$%
, $y$ and $z$ components, and $\sigma $ is the Pauli spin matrix. The $%
V_{soc}$ between $\left\vert 5p_{x}^{{}}\right\rangle $ and $\left\vert
5p_{y}^{{}}\right\rangle $ results in the splitting of states with opposite
spin configurations, while that between $\left\vert 5p_{z}^{{}}\right\rangle
$ and $\left\vert 5p_{x}^{{}}\right\rangle $ ($\left\vert
5p_{y}^{{}}\right\rangle $) leads to the splitting of states and an
interchange of spin configurations.

In a uniform perpendicular magnetic field, the $\mathbf{B}_{z}$-induced
periodical Peierls phases can modulate the hopping integral as\ $\gamma
_{oo^{\prime }}^{\mathbf{R}_{ij}}(\mathbf{B}_{z})=\gamma _{oo^{\prime }}^{%
\mathbf{R}_{ij}}\exp i(\frac{2\pi }{\Phi _{0}}\int_{r_{j}}^{r_{i}}\mathbf{A}(%
\mathbf{r})\cdot d\mathbf{r})$, where $\Phi _{0}$ ($hc/e$) is the flux
quantum. In the Landau gauge $\mathbf{A}=(0,B_{z}x,0)$, the unit cell
becomes an enlarged rectangle with $4R_{B}$ ($4\times 21665/B_{z}$) atoms
(Fig. 1(a)), where $R_{B}$ is the ratio of $\Phi _{0}$ versus magnetic flux
through each hexagon. The reduced Brillouin zone has an area of $4\pi ^{2}/%
\sqrt{3}a^{2}R_{B}.$ The magnetic Hamiltonian is built from the space
spanned by the $32R_{B}$ tight-binding functions $\left\{ \left\vert
A_{om}^{i}\right\rangle ;\left\vert B_{om}^{i}\right\rangle \left\vert
i=1,2,\cdots ,2R_{B};\text{ }o=5p_{x},5p_{y},5p_{z},5s;\text{ }m=\uparrow
,\downarrow \right. \right\} $, related to the \textit{i}th A and B atoms in
the rectangular unit cell. An electric field along the $z$-axis introduces a
potential energy $V_{z}/2$ ($-V_{z}/2$) to the site energy of the A (B)
sublattice. Due to the $sp^{3}$ hybridization, the hopping integrals of the
three nearest neighbors might have different signs, e.g., $\gamma _{xy}^{%
\mathbf{R}_{2}}=-\gamma _{xy}^{\mathbf{R}_{3}}$, which means the
corresponding matrix elements are complex for any wavevectors ($\mathbf{k}%
^{\prime }s$). In our calculations, the $32R_{B}\times 32R_{B}$ Hermitian
matrix is transformed into a band-like matrix, in which the base functions
are arranged in a specific sequence$^{23}$ $\left\{ \left\vert
A_{om}^{1}\right\rangle ,\left\vert B_{om}^{2R_{B}}\right\rangle ,\left\vert
B_{om}^{1}\right\rangle ,\left\vert A_{om}^{2R_{B}}\right\rangle ,\left\vert
A_{om}^{2}\right\rangle ,\left\vert B_{om}^{2R_{B}-1}\right\rangle
,\left\vert B_{om}^{2}\right\rangle ,\left\vert
A_{om}^{2R_{B}-1}\right\rangle \cdots \right\} $. Numerical diagonalization
for solving the eigenvalues ($E^{c,v\prime }s$) and eigenfunctions ($\Psi
^{c,v\prime }s$) is more efficient. The generalized tight-binding model
could be further employed to comprehend the magnetic quantization in other
layered systems with complex orbital bonding and spin configurations.

\bigskip

\bigskip

\bigskip

\bigskip \bigskip \centerline {\textbf {III. RESULTS AND DISCUSSION}}%
\bigskip \bigskip

Tinene has feature-rich energy bands, mainly owing to the critical SOC and $%
sp^{3}$ bonding. Two pairs of energy bands, out of a total of four, lie near
$E_{F}$. In this two pairs, each electronic state is doubly degenerate for
the spin degree of freedom. Without the SOC (black curves in Fig. 2(a)), one
pair of linear bands intersects at the K (K$^{\prime }$) point, and two
pairs of parabolic bands appear near the $\Gamma $ point because of the
strong $sp^{3}$ bonding. The former and the latter mainly originate from the
$5p_{z}$ and ($5p_{x},5p_{y}$) orbitals, respectively (Figs. 2(b) and 2(c)).
$V_{SOC}$ further induces the anti-crossing of energy\ bands and the
splitting of degenerate states (green curves). At the K (K$^{\prime }$)
point, the Dirac point is separated, and the doubly degenerate wavefunctions
of the first valence band are dominated by $\left\vert 5p_{z}^{B};\downarrow
\right\rangle $ and $\left\vert 5p_{z}^{A};\uparrow \right\rangle $ ($%
\left\vert 5p_{z}^{B};\uparrow \right\rangle $ and $\left\vert
5p_{z}^{A};\downarrow \right\rangle $), where only the first degenerate
state is shown (black curves in Fig. 2(b)). Along K$\rightarrow $M (K$%
^{\prime }\rightarrow \Gamma $), the contribution of $\left\vert
5p_{z}^{B};\downarrow \right\rangle $ ($\left\vert 5p_{z}^{B};\uparrow
\right\rangle $) quickly declines, while the opposite is true for that of $%
\left\vert 5p_{z}^{A};\downarrow \right\rangle $ ($\left\vert
5p_{z}^{A};\uparrow \right\rangle $). For the first conduction band, the $%
5p_{z}$-dependent wavefunction exhibits similar behavior under the
interchange of the two sublattices (Fig. 2(c)). These results imply that the
$\pi $- and $\pi ^{\ast }$-electronic states strongly depend on the spin
configurations and the two sublattices. On the other hand, in the vicinity
of the $\Gamma $ point, the state degeneracy is lifted by the $V_{SOC}$
between $5p_{x}$ and $5p_{y}$ orbitals, and one of the parabolic bands is
close to $E_{F}$. An indirect energy gap, which corresponds to the highest
occupied state at the $\Gamma $ point and the lowest unoccupied state at the
K point, is about $54$ meV. In addition, the anti-crossings along $\Gamma
\rightarrow $K$^{\prime }$ ($\Gamma \rightarrow $M) are attributed to the $%
V_{SOC}$ of the $5p_{z}$ and ($5p_{x},5p_{y}$) orbitals. $\left\vert \Psi
^{v_{1}}(\Gamma )\right\vert ^{2}$ has almost the same weight on the A and B
sublattices, but the spin-up state is much stronger than the spin-down
state. Along $\Gamma \rightarrow $K$^{\prime }$ ($\Gamma \rightarrow $M), $%
\left\vert \Psi ^{v_{1}}(\mathbf{k})\right\vert ^{2}$ is dominated by the
spin-up states (spin-up and spin-down states). That is to say, the spin
configurations are anisotropic. Tinene possesses special low-lying energy
bands quite different from the linear ones of graphene, silicene and
germanene. It therefore exhibits rich magnetic quantization. It should also
be noted that the tight-binding model calculations are consistent with the
first-principles ones.$^{15}$

\bigskip

Tinene has two groups of well-behaved low-lying LLs, in which the valence
and conduction LLs of each group are asymmetric about $E_{F}$ (Figs. 3(a)
and 3(c)). The first and second groups, respectively, correspond to the
magnetic quantization of the electronic states near the K (K$^{\prime }$)
and $\Gamma $ points. For each ($k_{x},k_{y}$), the LLs of the former are
eight-fold degenerate, except for the initial valence and conduction ones,
being attributed to the K and K$^{\prime }$ valleys, the mirror symmetry of $%
z=0$ plane, and the spin degree of freedom. At ($k_{x}=0,k_{y}=0$), the
degenerate states are localized at the $1/6,$ $2/6,$ $4/6$ and $5/6$
positions of the enlarged unit cell. The $2/6$ ($1/6$) states, analogous to
the $5/6$ ($4/6$) ones, originate from the electronic states in the K (K$%
^{\prime }$) valley. On the other hand, the LLs of the second group are
doubly degenerate, since there is only one \ $\Gamma $ valley, and the $%
V_{SOC}$ between the $5p_{x}$ and $5p_{y}$ orbitals leads to distinct
spin-dominated LL states (Fig. 3(d)). The two degenerate states are similar
but localized at the $0$ and $1/2$ positions of the enlarged unit cell. In
addition, the carrier density, which can be occupied in each localization-
and spin-distinct LL state, is the area of the reduced first Brillouin zone.

\bigskip

Each LL wavefunction is characterized by the subenvelope functions on the
different lattices with the $sp^{3}$ orbitals and spin configurations. The
subenvelope functions are well-behaved in their spatial distributions, with
a normal zero-point number and spatial symmetry or anti-symmetry about the
localization center. Furthermore, they are similar to those of a harmonic
oscillator. For the first group of LLs, the $1/6$ ($4/6$) states can be
regarded as the $2/6$ ($5/6$) states under the interchange of two
sublattices. The $2/6$ states, as shown in Fig. 3(b), are chosen to
illustrate the main features of the LL wavefunctions. Apparently, the
subenvelope functions of the $5p_{z}$ orbitals (black curves) dominate the
spatial distributions where the odd- and even-indexed A (B) sublattices have
a similar quantum mode. Since the wavefunctions of the first valence and
conduction LLs are mainly determined by the B sublattice, its zero-point
number can characterize a quantum number. The $n_{\text{K}}^{c,v}>0$ LLs are
composed of the spin-down and spin-up configurations. For the valence LLs
with small $n_{\text{K}}^{v}%
%TCIMACRO{\U{b4}}%
%BeginExpansion
{\acute{}}%
%EndExpansion
s$, B$_{o\downarrow }$ has a much larger weight compared to A$_{o\downarrow }
$. The difference between the weights declines with an increase of $n_{\text{%
K}}^{v}$ and becomes negligible at $n_{\text{K}}^{v}>20$. Also, the
conduction LLs behave similarly, except that the weight of A$_{o\downarrow }$
is stronger than that of B$_{o\downarrow }$. These facts reflect the
non-equivalence of the A and B sublattices in the vicinity of \ the K point.
The degenerate $\downarrow $- and $\uparrow $-dominated LL wavefunctions
resemble each other under the interchange of the weights of A and B
sublattices. Specifically, the $n_{\text{K,K}^{\prime }}^{c}=0$ ($n_{\text{%
K,K}^{\prime }}^{v}=0$ ) LLs with four-fold degeneracy are spin-polarized
because $V_{SOC}$ breaks the spin degeneracy. In addition, the K and K$%
^{\prime }$ points make the same contribution.

The second group is in sharp contrast to the first one regarding the main
features of the LL wavefunctions, including the orbital contributions,
weights of spin-up and spin-down components, and quantum modes on the A and
B sublattices. The spin-split LLs present almost identical $5p_{x}$- and $%
5p_{y}$-orbital subenvelope functions with the same zero-point number on
different sublattices (red and green curves in Fig. 3(d)), defined as the
quantum number $n_{\Gamma \uparrow \downarrow }^{c,v}$. For the $n_{\Gamma
\uparrow }^{v}$ ($n_{\Gamma \downarrow }^{v}$) LLs, the ratio of the $%
\uparrow $- to $\downarrow $-dependent components ($\downarrow $- to $%
\uparrow $-dependent ones) is about three. However, the $n_{\Gamma \uparrow
}^{c}$ and $n_{\Gamma \downarrow }^{c}$ LL wavefunctions are, respectively,
governed by the $\uparrow $- and $\downarrow $-dependent components, leading
to the larger LL splitting. As a result of the hexagonal lattice, the
nearest-neighbor atomic interactions near the $\Gamma $ point are roughly
proportional to $k^{2}$. The A and B sublattices present the same quantum
mode after magnetic quantization. However, the quantized electronic states
near the K or K$^{\prime }$ point are associated with the linear $k$%
-dependence of the nearest-neighbor interactions, so that the differcence of
quantum modes on the two sublattices is $\pm 1$.

The low-lying LLs present an unusual $B_{z}$-dependent energy spectrum, as
clearly shown in Fig. 4(a). The first- and second-group LLs frequently and
exclusively cross one another (black and red curves), since they have
different localization centers. The $B_{z}$-dependent energies directly
reflect the zero-field energy bands, especially for the linear and parabolic
bands. The former have roughly $\sqrt{B_{z}}$-dependent energies, except for
the $n_{\text{K,K}^{^{\prime }}}^{c,v}=0$ LLs, which have constant energies.
However, the $B_{z}$-dependence of the latter is approximately linear. With
the increment of $B_{z}$, the energy spacing between the $n_{\Gamma
\downarrow }^{c}$ and $n_{\Gamma \uparrow }^{c}$ LLs grows, arising from the
enhanced $V_{SOC}$ by the more localized LL wavefunctions. In addition, the
energy gap slightly increases because of the variation in the $n_{\Gamma
\downarrow }^{v}=0$ LL spectrum. The important differences between two
groups of LLs are clearly revealed in the density of states (DOS) (Fig.
4(b)). The single- and double-peak structures come from the first- and
second-group LLs, respectively. They appear at the different energy ranges,
with the former having larger peak spacings. The peak height is determined
by the number of degenerate states. The intensity of the single-peak
structures related to the $n_{\text{K,K}^{^{\prime }}}^{c,v}=0$ ($n_{\text{%
K,K}^{^{\prime }}}^{c,v}>0$) LLs is two (four) times higher than that of the
double-peak ones.

\bigskip

The magneto-electronic properties are diversified by a perpendicular
electric field. The $V_{z}$-dependent energy spectrum, shown in Fig. 5(a),
presents non-monotonous dispersion relations for the first group, but
monotonous ones for the second group. The K-valley- and K$^{\prime }$%
-valley-dependent LLs are further split by the destruction of the mirror
symmetry. The two splitting LLs have an opposite $V_{z}$-dpendence before $%
V_{z}$ reaches the critical potential ($V_{c}\sim 147$ meV) corresponding to
the linear bands intersecting at the K (K$^{\prime }$) point (blue curves in
the inset of Fig. 5(a)). With a further increase of $V_{z}$, all LLs move
gradually away from $E_{F}$. As to the second group, the weight of the B
sublattice is enhanced by $V_{z}$, and thus the valence LL energies are
reduced. The complicated $V_{z}$-dependences result in many LL crossings and
certain LL anti-crossings. In addition to the intergroup LL crossings, the
former come from intragroup LLs with distinct quantum modes, e.g., the
direct crossing of the $n_{\text{K}^{\prime }}^{v}=1$ and $2$ LLs (arrows in
Fig. 5(b)). Moreover, the intragroup LL anti-crossings only occur for the $%
n_{\text{K}}^{v}$ and $n_{\text{K}}^{v}\pm 1$ valence LLs, as well as the $%
n_{\text{K}^{\prime }}^{c}$ and $n_{\text{K}^{\prime }}^{c}\pm 1$ conduction
LLs (blue rectangles in Fig. 5(a)). The $V_{z}$-dependent energy spectrum
directly reflects the band structures at different $V_{z}^{^{\prime }}s$ in
the absence of a magnetic field.

There exist dramatic changes in the spatial distributions during the LL
anti-crossing, as illustrated by the $n_{\text{K}}^{v}=1$ and $2$ LLs in the
range of $20$ meV$\leq V_{z}\leq 60$ meV (purple and green triangles in Fig.
5(b)). At small $V_{z}^{^{\prime }}s$, the $5p_{z}$-orbital subenvelope
functions of the $n_{\text{K}}^{v}=1$ LL possess large and small weights on
the (A$_{o\downarrow }$, B$_{o\downarrow }$) and (A$_{o\uparrow }$, B$%
_{o\uparrow }$) sublattices, respectively (Fig. 5(c)). Although the ($%
5p_{x}^{{}},5p_{y}^{{}}$)-orbital subenvelope functions possess small
weights, they play an important role in the process of LL anti-crossings.
The intra-atomic $V_{SOC}$ between the $5p_{z}$ and ($%
5p_{x}^{{}},5p_{y}^{{}} $) orbitals induces a probability distribution
transfer between A$_{o\downarrow }$ and A$_{o\uparrow }$ (B$_{o\downarrow }$
and B$_{o\uparrow } $). Furthermore, the electric field leads to a similar
transfer between the A and B sublattices. By means of the cooperation
relationship of the electric field and $V_{SOC}$, the weights of the (A$%
_{o\downarrow }$, B$_{o\downarrow }$) and (A$_{o\uparrow }$, B$_{o\uparrow }$%
) sublattices become comparable when LL anti-crossings take place. On the
other hand, the $5p_{z}$-orbital subenvelope functions of the $n_{\text{K}%
}^{v}=2$ LL exhibit quantum modes identical to those of the $n_{\text{K}%
}^{v}=1$ LL, except for the interchange of the weights of spin-up and
spin-down configurations (Fig. 5(d)). At $V_{z}\sim 40$ meV, the $n_{\text{K}%
}^{v}=1$ and $2$ LLs are forbidden to have the same energy since their
subenvelope functions are almost identical. With a further increase of $%
V_{z} $, the relationship between the electric field and $V_{SOC}$ turns
into a competing one. As a result, the two LLs recover to the distinct
quantum modes at $V_{z}\geq 55$ meV. Similar anti-crossings can also be
observed in the conduction LLs. Additionally, in the coexistent LL crossing
and anti-crossing (Fig. 5(b)), two double peaks in DOS are changed into
three single peaks, or vice versa (not shown).

The $V_{z}$-dependent LL energy spectra of silicene, germanene and tinene
are quite different, as shown in Figs. 6(a) and 6(b) at $B_{z}=14$ T. The
low-lying second-group LLs are absent in the former two due to the weaker $%
sp^{3}$ bonding; that is, the low-energy magnetic quantization is mainly
determined by the $\pi $-electronic states near the K and K$^{^{\prime }}$
valleys. Their energy spacings, like in graphene, are proportional to the
Fermi velocity ($v_{F}=3bV_{pp\pi }/2$). Since silicene has the largest $%
v_{F}$ and the weakest $V_{SOC}$, neither the LL crossings nor the
anti-crossings are induced during a variation of $V_{z}$. However, germanene
presents intragroup LL crossings and anti-crossings similar to those of
tinene. The anti-crossings occur at larger electric fields and exist in a
wider range; they are attributed to the smaller $V_{SOC}$ (0.196 eV) and the
different $sp^{3}$ bondings.

STS measurements, in which the tunneling conductance (dI/dV) is
approximately proportional to DOS and directly reflects its special
structures, can provide an efficient way to identify the energy spectra
under external magnetic and electric fields. This method has been
successfully used to investigate the magneto-electronic energy spectra of
graphene-related systems, including the $\sqrt{B_{z}}$-dependent LL energies
in monolayer graphene,$^{27,28}$ the linear $B_{z}$-dependence in AB bilayer
graphene,$^{28,34}$ the coexistent $\sqrt{B_{z}}$- and $B_{z}$-dependences
in ABA trilayer graphene,$^{35}$ the absence of a simple $B_{z}$-dependence
in twisted bilayer graphene,$^{34}$ and the bulk and layer properties of the
Landau subbands in graphite.$^{36,37}$ Moreover, the electric-field-induced
energy gaps are confirmed in AB-bilayer and ABC-trilayer graphenes.$^{38-40}$
The main features of the magneto-electronic properties of monolayer
tinene---two groups of LLs, the $B_{z}$- and $V_{z}$-dependent splittings,
and the intragroup crossings and anti-crossings---can be further
investigated with STS. The STS measurements on the structure, energy range,
spacing and intensity of the prominent DOS peaks can verify the complex $%
sp^{3}$ bonding, the critical spin-orbital interaction, and the cooperative
or competitive relationship between $V_{SOC}$ and $V_{z}$.

\bigskip \bigskip \centerline {\textbf {IV. SUMMARY AND CONCLUSIONS}}%
\bigskip \bigskip

The novel magnetic quantization in monolayer tinene, being closely related
to the $sp^{3}$ bonding, spin-orbital coupling, and the influence of
magnetic and electric fields, is investigated by the generalized
tight-binding model. The rather large Hamiltonian matrix is built from the
tight-binding functions of the different sublattices, atomic orbitals and
spin configurations. It has complex elements due to the special relations
among the $sp^{3}$ orbitals and can be solved by using an exact
diagonalization method. The two groups of feature-rich LLs, which are,
respectively, dominated by the $5p_{z}$ orbitals and ($5p_{x}$,$5p_{y}$)
orbitals, are revealed near the Fermi level simultaneously. The important
differences between them lie in their spatial distributions, state
degeneracy, spin configurations, and $B_{z}$- and $V_{z}$-dependence. There
are approximate formulas in the B$_{z}$-dependent energy spectra, but not in
the V$_{z}$-dependent ones. The LL splittings in the first and second groups
are induced by the effects of electric and magnetic fields, respectively.
Specifically, only the $5p_{z}$-dominated group exhibits the LL
anti-crossings during the variation of $V_{z}$, arising from the cooperation
relationship between the electric field and spin-orbital coupling. These
unique magneto-electronic properties in tinene are absent in graphene,
silicene and germanene. The predicted magneto-electronic energy spectra
could be directly verified by STS measurements.

\bigskip

\bigskip

\centerline {\textbf {ACKNOWLEDGMENT}}

\bigskip

\bigskip

\noindent \textit{Acknowledgments.} This work was supported by the NSC of
Taiwan, under Grant No. NSC 102-2112-M-006-007-MY3.

\newpage

\noindent ~~~~$^\star$e-mail address: mflin@mail.ncku.edu.tw

\bigskip \vskip0.6 truecm

\noindent

\newpage

Figure captions

Figure 1. (a) The ($x$,$y$)-plane projection of monolayer tinene with a
rectangular unit cell in $B_{z}$ $\widehat{z}$, (b) the ($5p_{x}$,$5p_{y}$,$%
5p_{z}$,$5s$) orbitals, and (c) the $\pi $ and $\sigma $ bondings.

Figure 2. (a) The valence and conduction bands along the high symmetry
points (green curves) and those (black curves) without the spin-orbital
interactions. Their state probabilities due to various orbitals located at
two sublattices with spin-up and spin-down arrangements are, respectively,
displayed in (b) and (c).

Figure 3. As for the K point, (a) the LL energies and (b) the probabilities
of the subenvelope functions near the localization center in a rectangular
unit cell at $B_{z}$=14 T. Similarly, (c) and (d) correspond to the $\Gamma $
point.

Figure 4. (a) The magnetic-field-dependent energy spectra of the first and
second groups (black and red curves), and (b) density of states at $B_{z}$%
=14 T.

Figure 5. The gate-voltage-dependent energy spectrum at $B_{z}$=14 T, and
(b) the LL crossings and anti-crossing within a certain range of $E^{v}$, in
which the evolution of distribution probabilities of subenvelope functions
during the intragroup anti-crossing is revealed in (c) and (d). The Fermi
level is indicated by the dashed red curve. Also shown in the inset of (a)
is the band structure at $B_{z}$=0 and a critical $V_{z}$.

Figure 6. Same plot as Fig. 5(a), but shown for (a) silicene and (b)
gemanene.

\begin{figure}[p]
    \centering
    \includegraphics[width=0.8\textwidth]{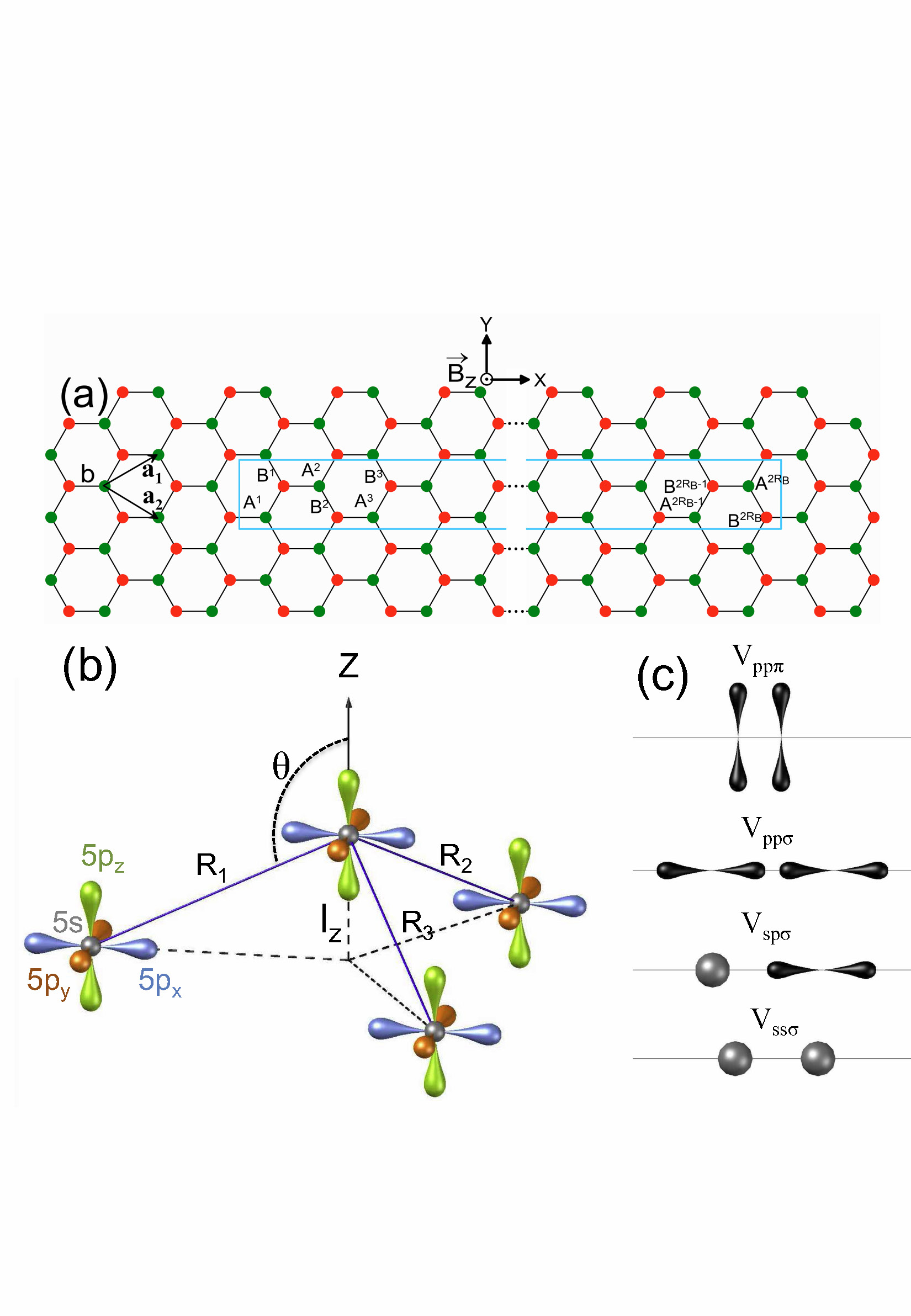}
    \caption{(a) The ($x$,$y$)-plane projection of monolayer tinene with a
rectangular unit cell in $B_{z}$ $\widehat{z}$, (b) the ($5p_{x}$,$5p_{y}$,$%
5p_{z}$,$5s$) orbitals, and (c) the $\pi $ and $\sigma $ bondings.}
    \label{figure:1}
\end{figure}

\begin{figure}[p]
    \centering
    \includegraphics[width=0.7\textwidth]{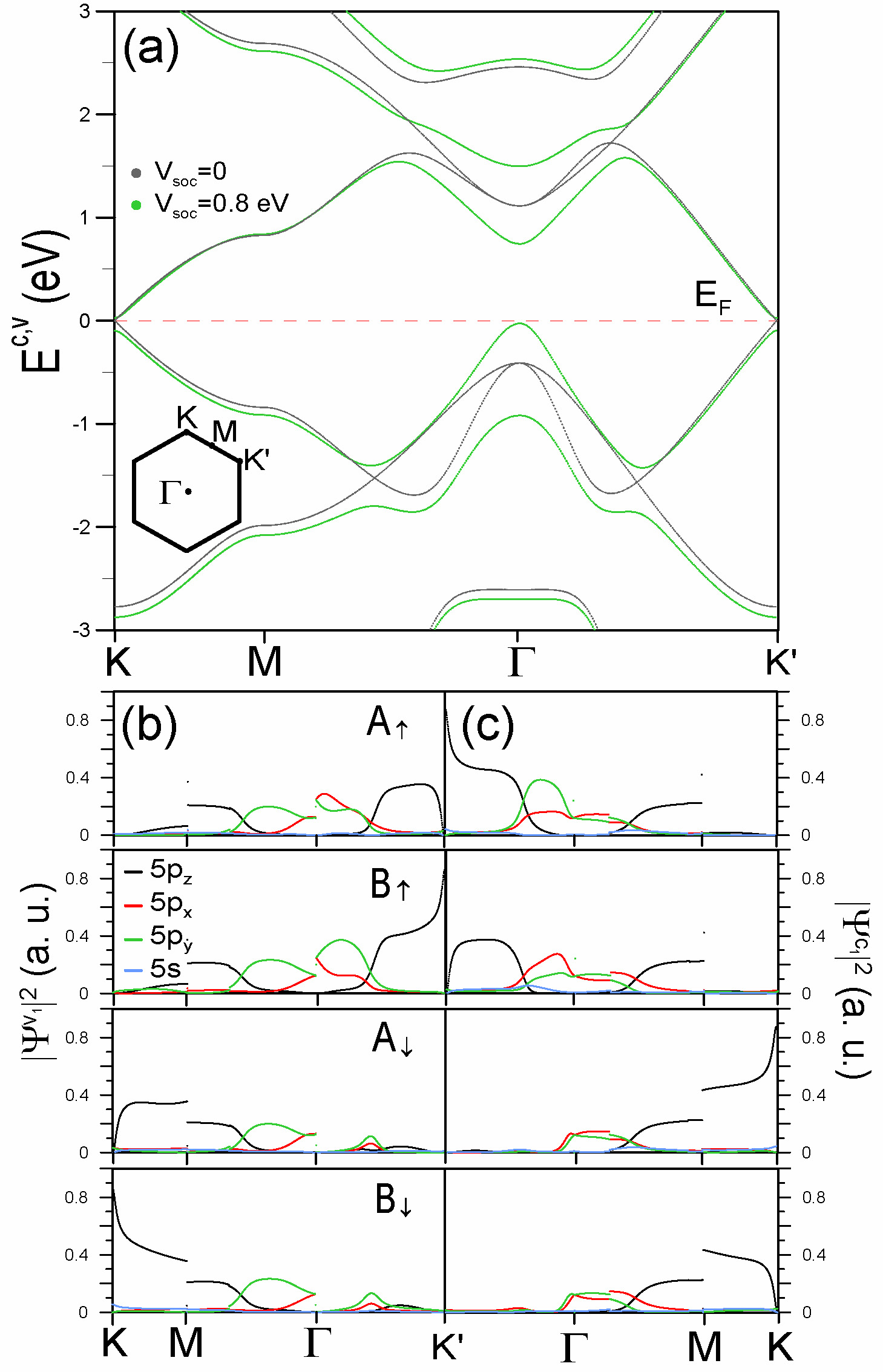}
    \caption{(a) The valence and conduction bands along the high symmetry
points (green curves) and those (black curves) without the spin-orbital
interactions. Their state probabilities due to various orbitals located at
two sublattices with spin-up and spin-down arrangements are, respectively,
displayed in (b) and (c).}
    \label{figure:2}
\end{figure}

\begin{figure}[p]
    \centering
    \includegraphics[width=0.8\textwidth]{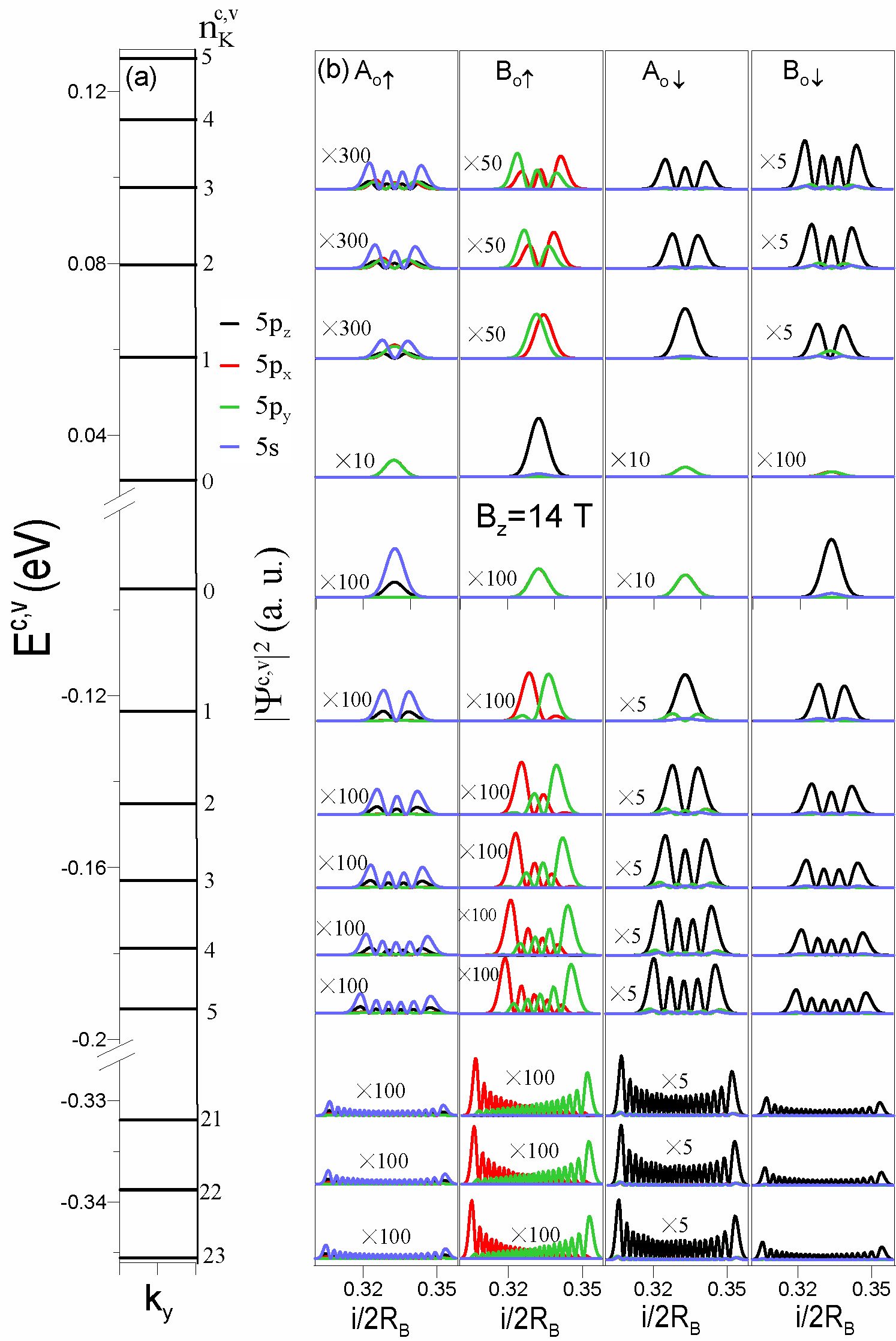}
    \caption{As for the K point, (a) the LL energies and (b) the probabilities
of the subenvelope functions near the localization center in a rectangular
unit cell at $B_{z}$=14 T. Similarly, (c) and (d) correspond to the $\Gamma $
point.}
    \label{figure:3a and 3b}
\end{figure}

\includegraphics[width=0.8\textwidth]{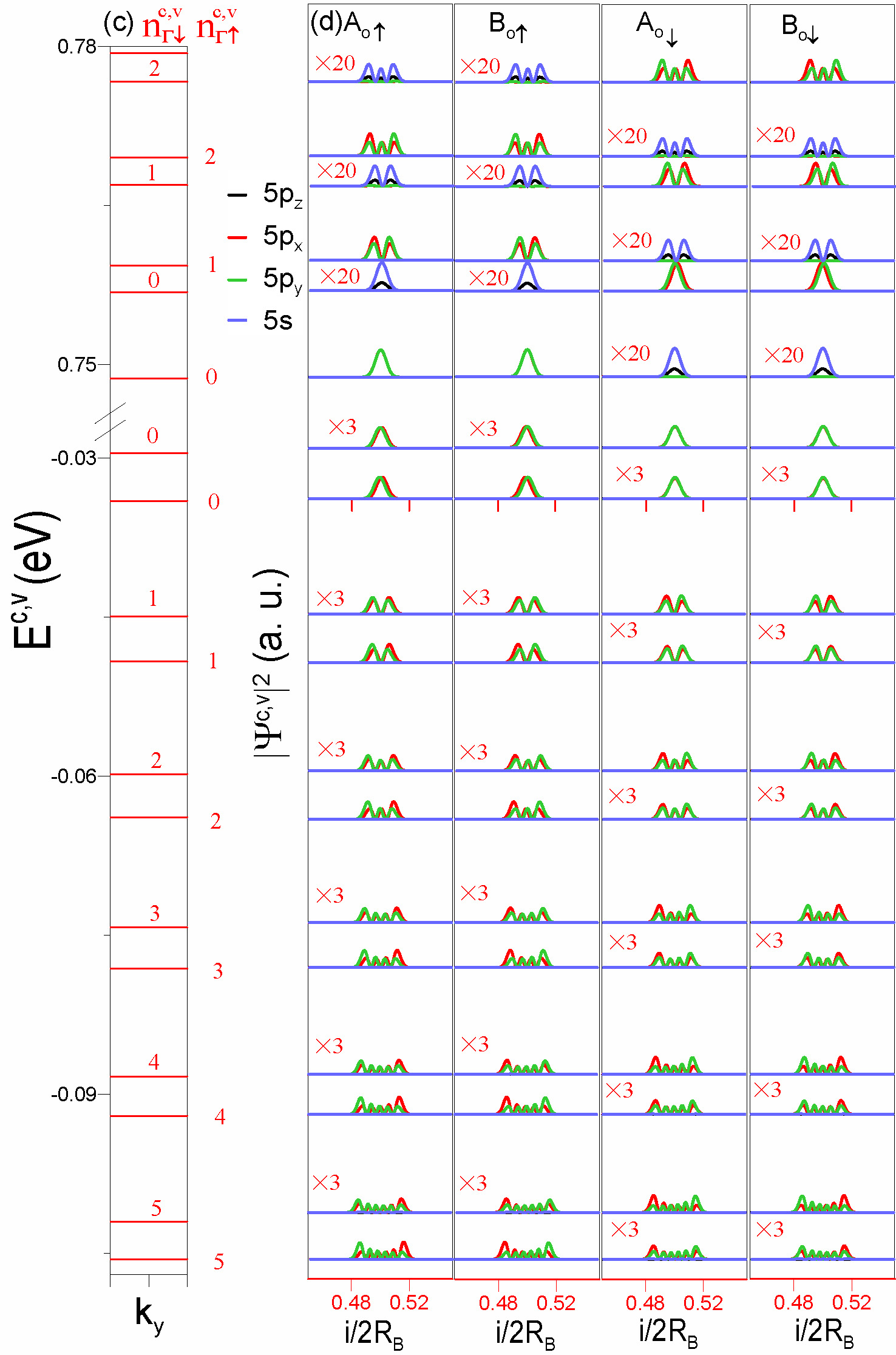}

\begin{figure}[p]
    \centering
    \includegraphics[width=0.8\textwidth]{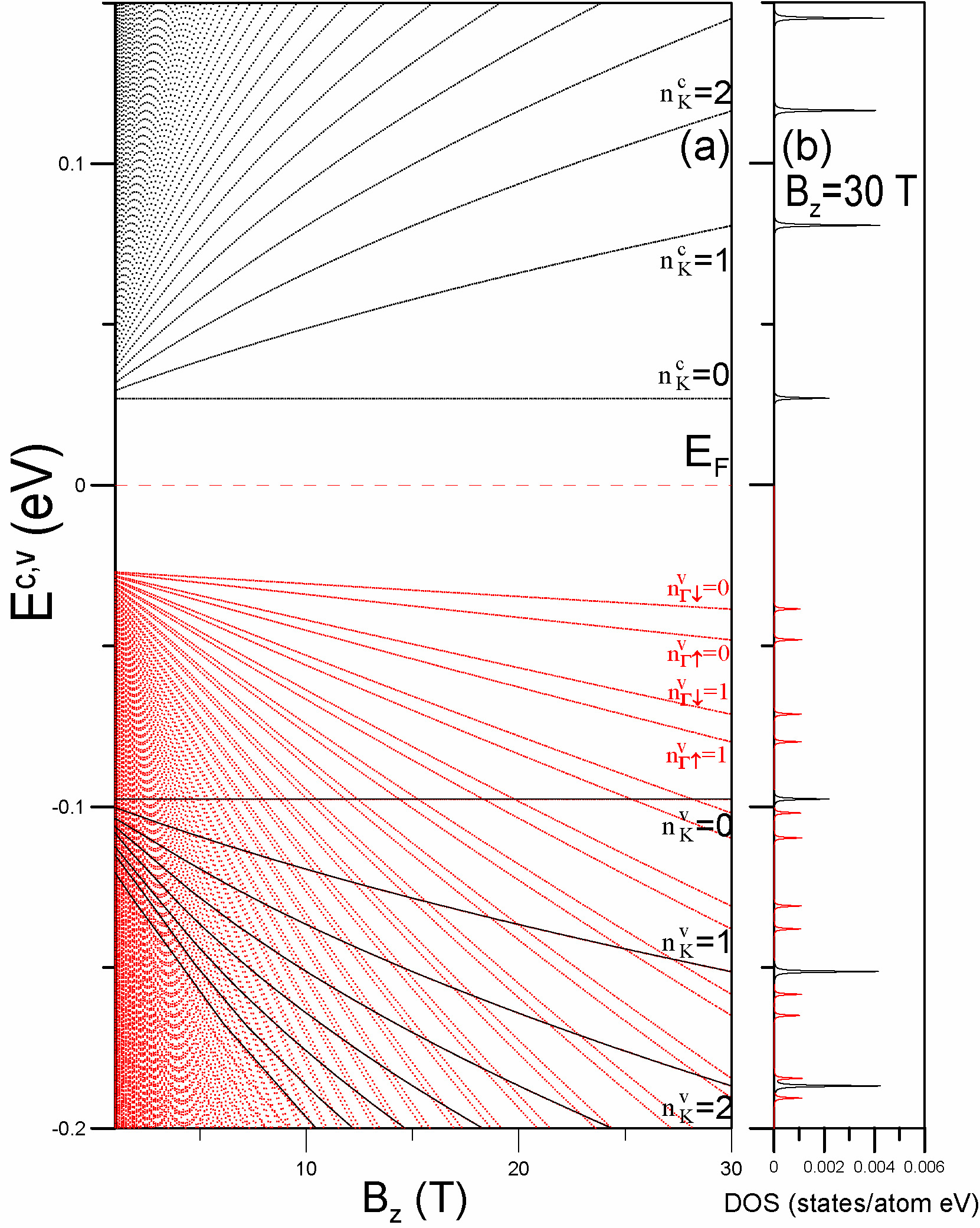}
    \caption{(a) The magnetic-field-dependent energy spectra of the first and
second groups (black and red curves), and (b) density of states at $B_{z}$%
=14 T.}
    \label{figure:4}
\end{figure}

\begin{figure}[p]
    \centering
    \includegraphics[width=0.8\textwidth]{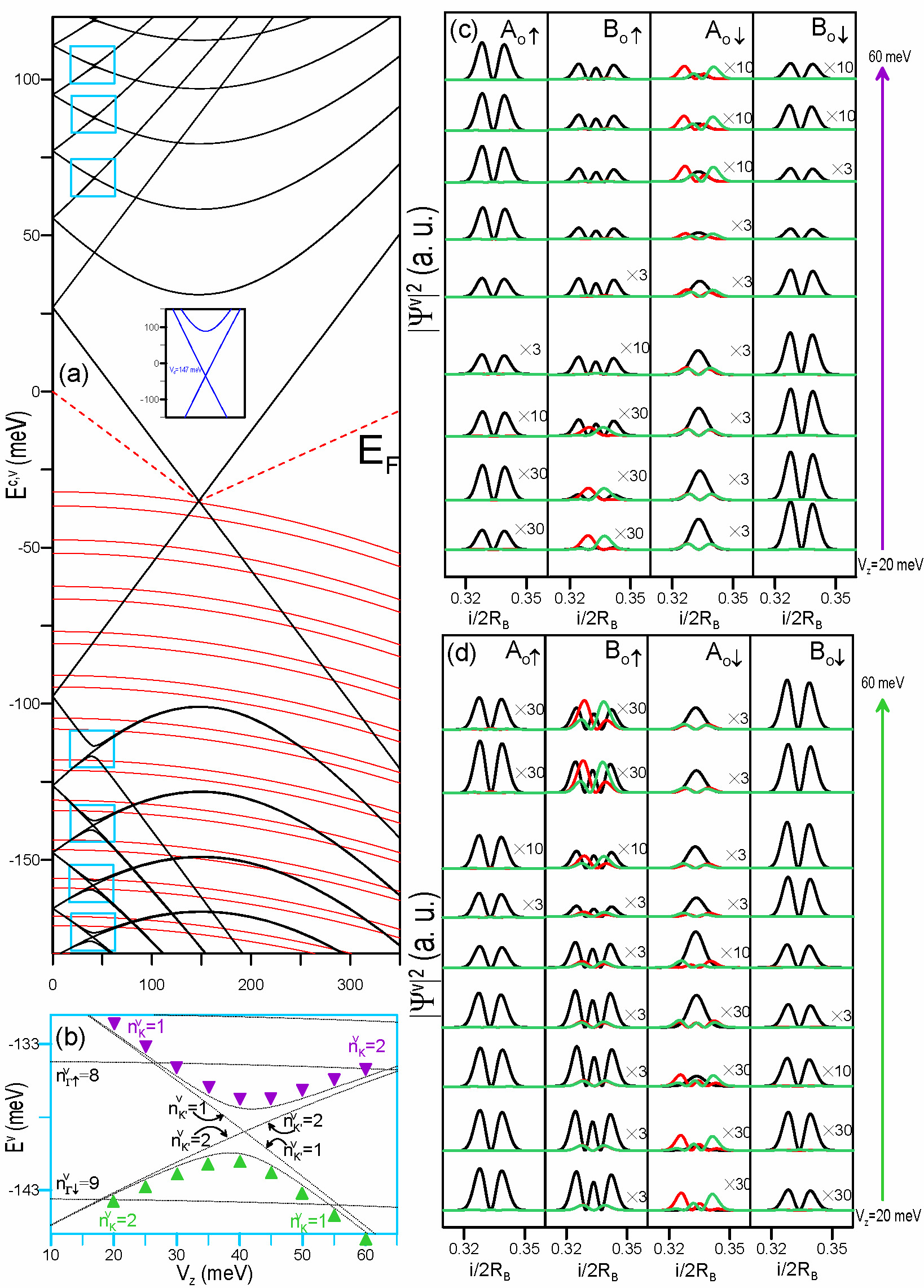}
    \caption{The gate-voltage-dependent energy spectrum at $B_{z}$=14 T, and
(b) the LL crossings and anti-crossing within a certain range of $E^{v}$, in
which the evolution of distribution probabilities of subenvelope functions
during the intragroup anti-crossing is revealed in (c) and (d). The Fermi
level is indicated by the dashed red curve. Also shown in the inset of (a)
is the band structure at $B_{z}$=0 and a critical $V_{z}$.}
    \label{figure:5}
\end{figure}

\begin{figure}[p]
    \centering
    \includegraphics[width=0.8\textwidth]{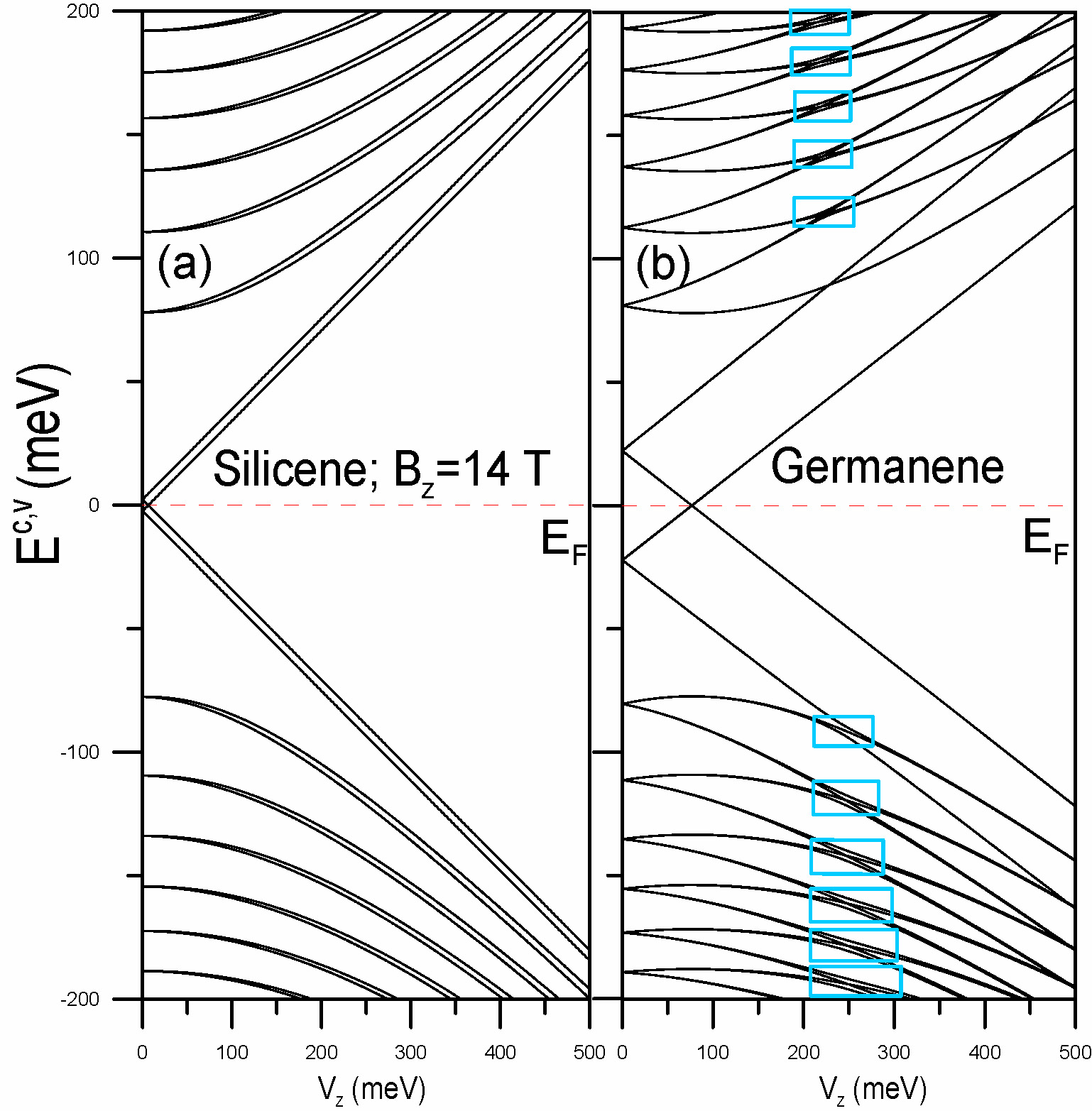}
    \caption{Same plot as Fig. 5(a), but shown for (a) silicene and (b)
gemanene.}
    \label{figure:6}
\end{figure}
\end{document}